\documentclass[aps,prc,
superscriptaddress,
preprint,
showpacs]{revtex4-1}
\usepackage{amsmath}
\usepackage{color}
\usepackage{graphicx}
\begin{document}
\title{New cluster approach on properties of 
  $^{8-11}$Be isotopes with isospin-dependent spin-orbit potential}

\author{Mengjiao Lyu} \email{mengjiao\_lyu@hotmail.com}
\affiliation{Department of Physics, Nanjing University, Nanjing
  210093, China} \affiliation{Research Center for Nuclear Physics
  (RCNP), Osaka University, Osaka 567-0047, Japan}

\author{Zhongzhou Ren} \email{zren@nju.edu.cn}
\affiliation{Department of Physics, Nanjing University, Nanjing
  210093, China} \affiliation{School of Physics Science
  and Engineering, Tongji University, Shanghai 200092, China}

\author{Hisashi Horiuchi} \email{horiuchi@rcnp.osaka-u.ac.jp}
\affiliation {Research Center for Nuclear Physics (RCNP), Osaka
  University, Osaka 567-0047, Japan} \affiliation {International
  Institute for Advanced Studies, Kizugawa 619-0225, Japan}

\author{Bo Zhou}
\affiliation{Faculty of Science, Hokkaido University, Sapporo
  060-0810, Japan}

\author{Yasuro Funaki} \affiliation{School of Physics and Nuclear
  Energy Engineering, Beihang University, Beijing 100191, China}

\author{\mbox{Gerd R\"{o}pke}} \affiliation{Institut f\"{u}r Physik,
  Universit\"{a}t Rostock, D-18051 Rostock, Germany}

\author{Peter Schuck} \affiliation{Institut de Physique Nucl\'{e}aire,
  Universit\'e Paris-Sud, IN2P3-CNRS, UMR 8608, F-91406, Orsay,
  France} \affiliation{Laboratoire de Physique et Mod\'elisation des
  Milieux Condens\'es, CNRS-UMR 5493, F-38042 Grenoble Cedex 9,
  France}

\author{Akihiro Tohsaki} \affiliation{Research Center for Nuclear
  Physics (RCNP), Osaka University, Osaka 567-0047, Japan}

\author{Chang Xu} \affiliation{Department of Physics, Nanjing University,
  Nanjing 210093, China}

\author{Taiichi Yamada} \affiliation{Laboratory of Physics, Kanto
  Gakuin University, Yokohama 236-8501, Japan}

\date{\today}

\begin{abstract}
  The nonlocalized clustering approach is generalized to $^{8-11}$Be
  isotopes with isospin dependent spin-orbit potential. A new form of
  the Tohsaki-Horiuchi-Schuck-R\"{o}pke (THSR) wave function is
  introduced to provide a correct description for the $\sigma$-binding
  neutron in $^{11}$Be. Systematic calculations for $^{8-11}$Be
  isotopes are performed and results fit well with experimental
  values. The low energy spectrum of $^{11}$Be is also obtained,
  especially the correct spin-parity $1/2^{+}$ is reproduced for the
  intruder ground state. The exotic neutron halo structure of
  $^{11}$Be is studied by calculations of root-mean-square radii and
  density distribution. We obtain a large spatial distribution for the
  last valence neutron of $^{11}$Be, which fits the phenomenological
  extracted value from experimental data. The spectroscopic
  factor is also calculated and discussed for the $1/2^+$ ground state of
  ${}^{11}$Be . 
\end{abstract}

\maketitle

The  studies of various spin-orbit interactions illuminate important
effects and mechanisms in different fields of physics.  Especially, a
large spin-orbit interaction is introduced in the development of the
nuclear shell model to reproduce the magic numbers such as 28, 50, 82,
and 126 for the $\beta$-stable nuclei \cite{Mayer1950}.  In recent
investigations, the spin-orbit coupling induces anomalous Hall effects
in tunnel junction or in nonmagnetic metals
\cite{Zhang2016,Matos2015}. The spin-orbit coupling also plays a key
role in various models for the understanding of topological insulators
\cite{Groth2009,Wang2010}.  Other prominent spin-orbit coupling
effects include the fine structure in atomic spectra and
magnetocrystalline anisotropy for materials \cite{Cullity2011,Griffiths2005}.

For neutron-rich nuclei which are away from the $\beta$-stable line,
spin-orbit couplings provide large contributions for the binding
 of the valence neutrons \cite{Itagaki2000,Enyo2002}.
Experimentally, the neutron-rich nucleus $^{11}$Be is famous for its
exotic properties including neutron halo structure and abnormal
positive ground state parity \cite{Chen2016,Aumann2000,Winfield2001,
Fukuda2004,Otsuka1993,Brown2001}. The key role of
large spin-orbit interaction has been discussed in some pioneer works
including studies with antisymmetrized molecular dynamics and shell
model calculations \cite{Enyo2002,Li2014,So2016,Otsuka1993}.  The dominance of
clustering effects in many Be isotopes including $^{11}$Be are also
confirmed in theoretical and experimental studies \cite{Funaki2002,
  Lyu2015,Lyu2016}. Thus, a new clustering approach with spin-orbit
interaction is essential for a valid description of the structure and
dynamics of the Be isotopes.

In recent years, the clustering effects and the halo structures in
nuclei have been extensively studied with various approaches
\cite{Tohsaki2017,Tohsaki2001, Funaki2002, Yamada2005, Zhou2012, Zhou2013,
  Zhou2014, Suhara2014,Lyu2015,Lyu2016, Xu2006, Yren2012, Ma2014,Spieker2015,
  Ye2014,Denisov2015,Lovas1998,Thompson1996,Rodriguez2010}. Especially,
the Tohsaki-Horiuchi-Schuck-R\"{o}pke (THSR) wave function with
intrinsic cluster degree of freedom is proposed and  underlies the new
nonlocalized concept for cluster dynamics \cite{Tohsaki2001,
  Funaki2002, Yamada2005, Zhou2012, Zhou2013, Zhou2014,
  Suhara2014}. Then in our previous studies, valence neutrons with
spin-orbit interaction  are treated in this model, what
successfully reproduces the physical properties of $^{9-10}$Be isotopes
and illustrates the dynamics of $\alpha$-clusters and valence
neutrons \cite{Lyu2015,Lyu2016}.

In this work, a new improved form of the THSR wave function is proposed
for the nucleus $^{11}$Be. With a new
introduced isospin-dependent spin-orbit interaction strength, we
perform the first systematic calculations for the $^{8-11}$Be isotopes
with the THSR wave function. The calculated physical properties 
agree well with
experimental results. Exotic properties of $^{11}$Be such
as the inversion of two lowest states and the large halo structure in
the ground state are also well described.

The THSR wave function of ${}^{11}$Be nucleus is formulated in the
form of creation operators, as
\begin{equation}
\label{eq:indwf}
  \left| \Phi (\text{Be}) \right\rangle
  =(C_{\alpha}^{\dagger})^{2}
   c_{n}^{\dagger}(9)
   c_{n}^{\dagger}(10)
   c_{n}^{\dagger}(11)
\left| { \bf  \rm vac} \right\rangle,
\end{equation}
where $C_{\alpha}^{\dagger}$ and $c_{n}^{\dagger}$ are creation
operators of $\alpha$-clusters and valence neutrons, respectively.
For the two $\alpha$-clusters in the $^{8}$Be core, 
the $\alpha$-creators $C_{\alpha}^{\dagger}$ are written as
\begin{equation}
  \label{eq:alpha-creator}
  \begin{split}
  C_{\alpha}^{\dagger}&=\int d\mathbf{R}
    \exp (-\frac{R_{x}^{2}
           +R_{y}^{2}}{\beta_{\alpha,xy}^{2}}
    -\frac{R_{z}^{2}}{\beta_{\alpha,z}^{2}})\int d\mathbf{r}_{1}
      \cdots d\mathbf{r}_{4}    \\
  &\times \psi(\mathbf{r}_{1}-\mathbf{R})
      a_{\sigma_{1},\tau_{1}}^{\dagger}(\mathbf{r}_{1})
    \cdots \psi(\mathbf{r}_{4}-\mathbf{R})
      a_{\sigma_{4},\tau_{4}}^{\dagger}(\mathbf{r}_{4}),
  \end{split}
\end{equation}
where $\mathbf{R}$ is the generate coordinate of the $\alpha$-cluster
and $a_{\sigma,\tau}^{\dagger}(\mathbf{r}_i)$ creates the $i$-th
nucleon with spin $\sigma$ and isospin $\tau$ at position
$\mathbf{r}_i$.
$\psi(\mathbf{r}) =(\pi b^{2})^{-3/4} \exp(-r^{2}/2b^{2})$ is the
single nucleon wave function. Parameter $b$ is chosen to be $b=1.35$
in this Gaussian function to optimize the binding energy of
$\alpha$-cluster. $\beta_{\alpha,xy}$ and $\beta_{\alpha,z}$ are
deformation parameters for the nonlocalized motion of two
$\alpha$-clusters.

Generally, the neutron creators $c_{n}^{\dagger}$ for the $i$-th
valence neutron can be written in the following form as
\begin{equation}
  \label{eq:neutron-creator}
  \begin{split}
   c_{n}^{\dagger}(i)&=\int d\mathbf{R}_{i}
    \exp (-\frac{R_{i,x}^{2}
           +R_{i,y}^{2}}{\beta_{i,xy}^{2}}
          -\frac{R_{i,z}^{2}}{\beta_{i,z}^{2}})
    \int d\mathbf{r}_{i}    \\
  &\times (\pi b^{2})^{-3/4}
  f(\mathbf{R}_{i},\cdots)
  e^{-(\mathbf{r}_{i}-\mathbf{R}_{i})^{2}/(2b^{2})}
    a_{\sigma}^{\dagger}(\mathbf{r}_{i}).
  \end{split}
\end{equation}
Here $\sigma$ is the spin of the valence neutron and
$a_{\sigma}^{\dagger}(\mathbf{r}_i)$ creates the $i$-th neutron with
spin $\sigma$ at position $\mathbf{r}_i$.  $\beta_{i,xy}$ and
$\beta_{i,z}$ are also  nonlocalization parameters for the $i$-th
valence neutron  in a deformed orbit.  This expression has a similar 
form as the $\alpha$-cluster creators in Eq.~\ref{eq:alpha-creator} 
except for the introduction of extra factor $f(\mathbf{R}_{i},\cdots)$.

For creators $c_{n}^{\dagger}(9)$ and $c_{n}^{\dagger}(10)$ of the first 
two valence neutrons in $^{11}$Be, we adopt the extra factor as
$f(\mathbf{R}_{9,10},\cdots)=e^{im\phi_{\mathbf{R}_{9,10}}}$, where
parameter $m=1$ with spin $\sigma=|\uparrow \rangle$ and $m=-1$ with
spin $\sigma=|\downarrow \rangle$. These two extra factors ensure
negative parity for the each single neutron wave function and provide
good description for their $\pi$-binding structure as we discussed in
our previous works.

\begin{figure}[htbp]
  \centering
  \includegraphics[width=0.55\textwidth]{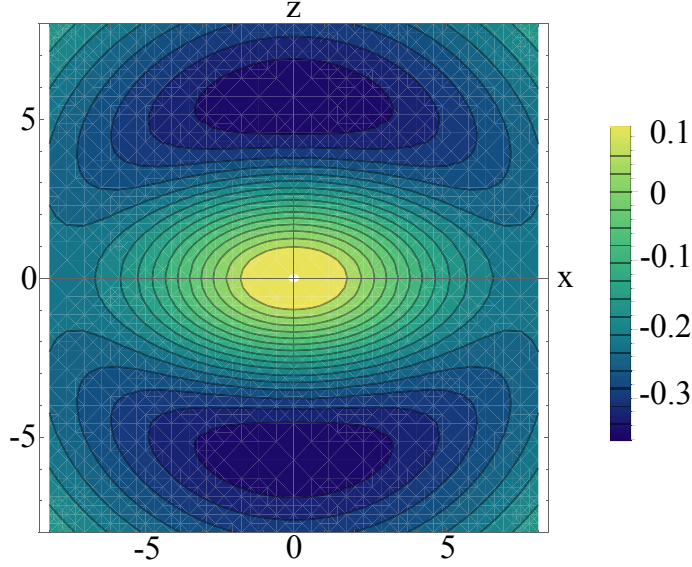}
  \caption{\label{fig:sigma-wf}The $y=0$ cross section of the single
    nucleon wave function
    $\phi_{n}(\mathbf{r})$ for the last valence neutron in the
    $1/2^{+}$ ground state of $^{11}$Be. The yellow color denotes
    positive values of $\phi_{n}(\mathbf{r})$. The blue color denotes
    negative values of $\phi_{n}(\mathbf{r})$. The green color denotes
    values close to zero of $\phi_{n}(\mathbf{r})$.  Parameters are
     chosen as $b=1.35$ fm, $\beta_{11,xy}=\beta_{11,z}=10$ fm, $a=3$
    fm, $\gamma=2$ and $|\mathbf{R}_{1}-\mathbf{R}_{2}|=1.3$ fm.}
\end{figure}

For the last valence neutron of the $^{11}$Be nucleus, we adopt
different creators for the $1/2^{+}$ and $1/2^{-}$
rotational bands. For the positive parity
states, we propose a new creation operator as
\begin{equation}
  \label{eq:sigma-creator}
  \begin{split}
  c_{n}^{\dagger}(11)&=\int d \mathbf{R}_{11}
  \exp \left(-\frac{R_{11,x}^{2}+R_{11,y}^{2}}{\beta_{11,xy}^{2}}
	-\frac{R_{11,z}^{2}}{\beta_{11,z}^{2}}\right)\\
&\times[
  \mathrm{e}^{ -\frac{R_{11,x}^{2}+R_{11,y}^{2}}{\gamma a^{2}}
	-\frac{R_{11,z}^{2}}{a^{2}}}
- \mathrm{e}^{-(\mathbf{R}_{1}-\mathbf{R}_{2})^{2}/(2a)^{2}
} ]\\
&\times\int d\mathbf{r}_{11}
     (\pi b^{2})^{-3/4}
  e^{-(\mathbf{r}_{11}-\mathbf{R}_{11})^{2}/(2b^{2})}
    a_{\uparrow}^{\dagger}(\mathbf{r}_{11}).
  \end{split}
\end{equation}
In this neutron creator, a new extra factor
\begin{equation}
  \label{eq:sigma-factor}
\begin{split}
f(\mathbf{R}_{11},\cdots)=&
  \exp \left(-\frac{R_{11,x}^{2}+R_{11,y}^{2}}{\gamma a^{2}}
	-\frac{R_{11,z}^{2}}{a^{2}}\right)\\
&- \exp \left(-(\mathbf{R}_{1}-\mathbf{R}_{2})^{2}/(2a)^{2}
\right)
\end{split}
\end{equation}
is introduced for a correct description of the nodal structure of the
$\sigma$-orbit wave function for the last valence neutron. The
generator coordinates $\mathbf{R}_{1}$ and $\mathbf{R}_{2}$ of two
$\alpha$-clusters are also included in this creator to describe the
correlation between the valence neutron and $\alpha$-clusters.  We
provide the analytical expression of the single nucleon wave function
$\phi_{n}(\mathbf{r})=\langle \mathbf{r} | c_{n}^{+} (i=11)|
\mathrm{vac} \rangle$ for the last valence neutron in the $1/2^{+}$
ground state of $^{11}$Be, as
\begin{equation}
  \label{eq:sigma-wf}
  \begin{split}
    &\phi_{n}(\mathbf{r})=\prod_{\xi=x,y,z}\Bigg \{
    \sqrt{\frac{1}{b^{2}}+\frac{2}{B_{\xi}^{2}}}
        \exp\left(-\frac{r_{\xi}^{2}}{2b^{2}+B_{\xi}^{2}}\right)\\
   &\,-\sqrt{\frac{1}{b^{2}}+\frac{2}{\beta_{\xi}^{2}}}
        \exp\left(-\frac{(\mathbf{R}_{1}-\mathbf{R}_{2})^{2}}{(2a)^{2}}
        -\frac{r_{\xi}^{2}}{2b^{2}+\beta_{\xi}^{2}}\right)
\Bigg \}.
  \end{split}
\end{equation}
Here, the subscript $\xi$  stands for  each of the $x,y$ and $z$  coordinates.
$1/B_{\xi}^{2}=1/\beta_{\xi}^{2}+1/(\gamma a)^{2}$ for $x$ and $y$
 coordinates and $1/B_{\xi}^{2}=1/\beta_{\xi}^{2}+1/a^{2}$ for the $z$
 coordinate.  In Fig.~\ref{fig:sigma-wf}, we show the $y=0$ cross
section for this single neutron wave function. An ellipsoidal  nodal
surface is clearly demonstrated in this figure  with green color. Along the $z$-axis,
the  nodes  with green color are physically the most  likely locations for
$\alpha$-clusters.

For the $1/2^-$ rotational band of ${}^{11}$Be with negative parity,
 we formulate $c_{n}^{\dagger}(11)$ with similar form as $c_{n}^{\dagger}(9)$
 and $c_{n}^{\dagger}(10)$ but with antiparallel spin-orbit coupling.

The Hamiltonian of the $^{11}$Be system can be written as
\begin{equation}\label{hamiltonian}
  H=\sum_{i=1}^{11} T_i-T_{c.m.} +\sum_{i<j}^{11}V^N_{ij}
    +\sum_{i<j}^{11}V^C_{ij} +\sum_{i<j}^{11}V^{ls}_{ij},
\end{equation}
where $T_{c.m.}$ is the kinetic energy of the center-of-mass
motion. Volkov No. 2  is used as the central force of
the nucleon-nucleon potential,
\begin{equation}\label{eq:nn-central}
  V^N_{ij}=\{V_1 e^{-\alpha_1 r^2_{ij}}-V_2 e^{-\alpha_2 r^2_{ij}}\}
  \{ W - M \hat P_\sigma \hat P_\tau \ + B \hat P_{\sigma} - H \hat P_{\tau}\},
	\end{equation}
where $M=0.6$, $W=0.4$ and $B=H=0.125$. Other parameters are
$V_{1}=-60.650$ MeV, $V_{2}=61.140$ MeV, $\alpha_{1}=0.309$
fm${}^{-2}$, and $\alpha_{2}=0.980$ fm${}^{-2}$.  The G3RS (Gaussian
soft core potential with three ranges) term is taken as the two-body
type spin-orbit interaction,
\begin{equation}\label{vc}
V^{ls}_{ij}=V^{ls}_0\{
           e^{-\alpha_1 r^2_{ij}}-e^{-\alpha_2 r^2_{ij}}
           \} \mathbf{L}\cdot\mathbf{S} \hat{P}_{31},
\end{equation}
where $\hat P_{31}$ projects the two-body system into triplet odd
state.  An isospin-dependent spin-orbit interaction strength is
adopted in our calculations, as
\begin{equation}\label{v0}
V^{ls}_0=2400+200\times\mathrm{|N-Z|\,\,\,\,(MeV) }
\end{equation}
Other parameters in $V^{ls}_{ij}$ are $\alpha_{1}$=5.00 fm${}^{-2}$,
and $\alpha_{2}$=2.778 fm${}^{-2}$.

\begin{figure}[htbp]
  \centering
  \includegraphics[width=0.55\textwidth]{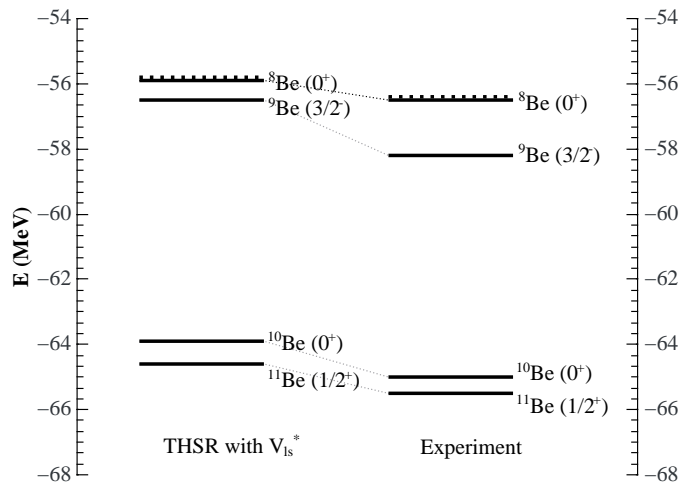}
  \caption{\label{fig:be-gs}The ground state energies of $^{8-11}$Be
    isotopes. ``THSR with $V_{ls}^{*}$" denotes calculated results
    with the THSR wave function and the modified spin-orbit
    interaction strength. ``Experiment" denotes the experimental
    values from Ref.~\cite{Wang2012}. The dashed lines indicate the
    corresponding $\alpha+\alpha+n+\cdots$ threshold.}
\end{figure}

\begin{figure}[htbp]
  \centering
  \includegraphics[width=0.55\textwidth]{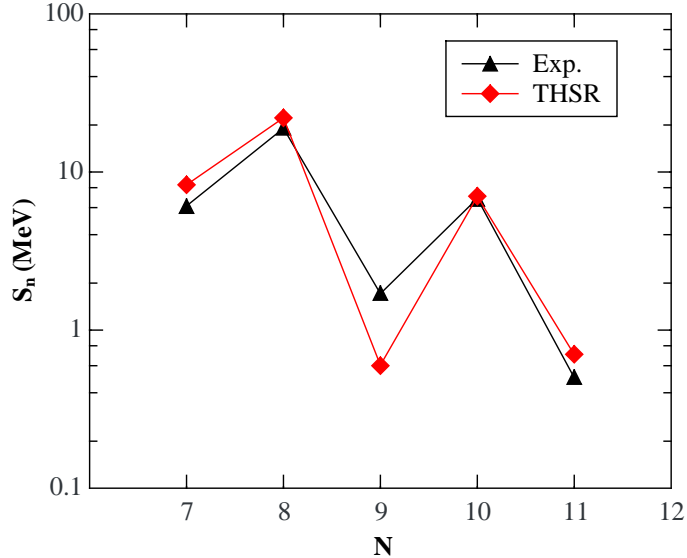}
  \caption{\label{fig:be-sn}The one-neutron separation energies
    $S_{n}$ of $^{6-11}$Be isotopes. ``THSR" denotes calculated
    results with the THSR wave function and ``Exp" denotes the
    experimental values from Ref.~\cite{Wang2012}.}
\end{figure}

We perform systematic calculations for the binding energies of
$^{8-11}$Be isotopes. The results are listed in Table
\ref{table:be-gs} and shown in Fig.~\ref{fig:be-gs}. The binding
energies calculated with the THSR wave function are found to be in
good agreement with corresponding experimental values. The $0^{+}$
ground state of $^{8}$Be, which is closely below the two $\alpha$
threshold, is correctly reproduced. The one neutron separation
energies $S_{n}$ of Be isotopes, as displayed in Fig.~\ref{fig:be-sn}, 
also fit corresponding experimental data.

\begin{table}[htbp]
\begin{center}
  \caption{\label{table:be-gs}Calculated results of the ground states
    of Be isotopes. ``E (THSR)'' denotes results calculated with the
    THSR wave function. ``$V_{ls}^{*}$'' denotes corresponding
    spin-orbit interaction strength. ``E (Exp)'' denotes experimental
    values adopted from Ref.~\cite{Wang2012}. All units are in MeV.  }
\begin{tabular}{l c c c c}
\hline
\hline
                 &E (THSR)  &E (Exp) \cite{Wang2012}  &$V_{ls}^{*}$\\
\hline
${}^{11}$Be$(1/2^{+})$ &-64.6 &-65.5 &3000\\
${}^{10}$Be$(0^{+})$ &-63.9 &-65.0 &2800\\
${}^{9}$Be$(3/2^{-})$ &-56.5 &-58.2 &2600\\
${}^{8}$Be$(0^{+})$ &-55.9 &-56.5 &2400\\
\hline
\hline
\end{tabular}
\end{center}
\end{table}

\begin{figure}[htbp]
  \centering
  \includegraphics[width=0.55\textwidth]{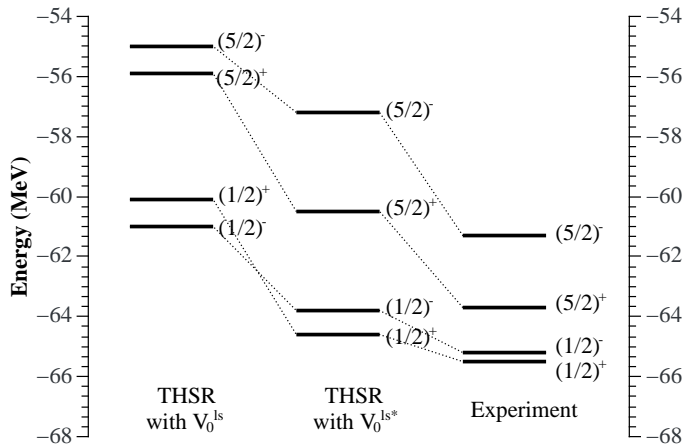}
  \caption{\label{fig:11be-band}The energy spectrum for the low lying
    states of $^{11}$Be. ``THSR with $V_{0}^{ls}$" denotes calculated
    results with fixed spin-orbit interaction strength
    $V_{0}^{ls}=2000$ MeV. ``THSR with $V_{0}^{ls*}$" denotes calculated
    results with the modified spin-orbit interaction strength
    $V_{0}^{ls*}=3000$ MeV. ``Experiment" denotes the experimental values from
    Ref.~\cite{Kelley2012}.}
\end{figure}

For the intruder ground state of $^{11}$Be, the exotic positive parity
is also correctly reproduced. As shown in Fig.~\ref{fig:11be-band},
the spin-orbit strength plays a key role for the level order for the
lowest $1/2^{+}$ and $1/2^{-}$ states.  It is shown that with the
weaker $V_{0}^{ls}=2000$ MeV strength, a negative parity is obtained
for the ground state. With the new strength $V_{0}^{ls*}=3000$ MeV,
which is 1000 MeV larger, the level order of the $1/2^{+}$ and
$1/2^{-}$ states is inverted. This inversion is due to that the
spin-orbit coupling in the $1/2^{+}$ ground state is larger than the
one in the $1/2^{-}$ state, where the last valence neutron has
antiparallel spin-orbit coupling.  We also see great improvement for
the low lying rotational band in Fig.~\ref{fig:11be-band},
which fits the experimental values reasonably well.
{ However, despite the correct ordering the spectrum stays quite a
bit too diluted compared with experiment. In the  phenomenological
particle-vibration coupling (PVC) model of Vinh Mau \cite{Vinhmau1995}, the spreading 
of the states is much reduced. This is probably due to the fact that 
experimental values of BE2 and other transition values are taken as 
input. Nevertheless, this hints to the fact that an explicit account 
of some low lying collective states of the ${}^{10}$Be core is needed. 
In our approach there is no adjustable parameter besides the increased 
spin-orbit strength but this is also the case in Ref.~\cite{Vinhmau1995}.}

The $^{11}$Be is also well known for its neutron halo structure. This
exotic property is studied with our THSR wave function by calculating
the root-mean-square radii for the ground state of $^{11}$Be, as
listed in Table \ref{table:be-rms}. It is found that the neutrons have
a much larger RMS radius than the protons in the ground state
$^{11}$Be. This indicates that the $1/2^{+}$ ground state of $^{11}$Be
has a neutron halo structure, which is consistent with the
experimental observations. In our calculations, the large difference
between neutron and proton radii comes mainly from the large spread of
the last valence neutron occupying the $\sigma$-orbit. In fact, a
giant RMS radius 8.94 fm is obtained for the last valence neutron,
which agrees with the giant value 7.98 fm extracted phenomenologically
from experimental data \cite{Tanihata2013}.
\begin{table}[htbp]
\begin{center}
  \caption{\label{table:be-rms}Matter, proton and neutron
    root-mean-square (RMS) radii of the ground state of $^{11}$Be
    denoted by $\langle r_{m}^{2}\rangle^{1/2} $,
    $\langle r_{p}^{2}\rangle^{1/2} $, and
    $\langle r_{n}^{2}\rangle^{1/2}$ respectively.
    $\langle r_{n,\sigma}^{2}\rangle^{1/2} $ denotes the RMS radius of
    the last valence neutron. Experimental values are adopted from
    Ref.~\cite{Tanihata2013}. All units are in fm.  }
\begin{tabular}{l c c c c c}
\hline
\hline
      &$\langle r_{m}^{2}\rangle^{1/2}  $
        &$\langle r_{p}^{2}\rangle^{1/2} $ 
          &$\langle r_{n}^{2}\rangle^{1/2}$  
            &$\langle r_{n,\sigma}^{2}\rangle^{1/2} $ \\
\hline
THSR &3.73 &2.42 &4.25 &8.94\\
Exp \cite{Tanihata2013} &2.91 &2.36 &3.09 &7.98\\
\hline
\hline
\end{tabular}
\end{center}
\end{table}

\begin{figure}[htbp]
  \centering
  \includegraphics[width=0.55\textwidth]{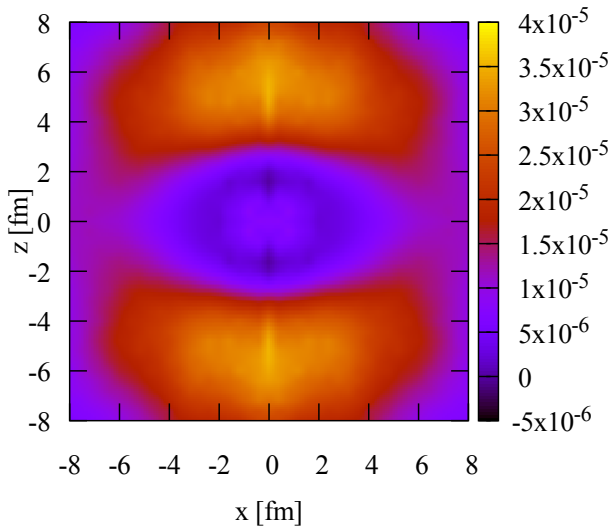}
  \caption{\label{fig:sigma-dens}Density distribution of
    valence neutrons occupying the $\sigma$-orbit in the
    intrinsic $1/2^{+}$ state of $^{11}$Be.  The color scale of each
    point in the figure is proportional to the nucleon density
    in the $x-z$ plane of the $y=0$ cross section. The unit of
    the density is fm$^{-3}$.  }
\end{figure}

The density distribution is shown in Fig.~\ref{fig:sigma-dens} to
illustrate the structure of the $1/2^{+}$ ground state and the giant
extension of the last valence neutron. To obtain the density distribution,
we first rewrite the intrinsic wave function $|\Psi\rangle$ of $^{11}$Be as
\begin{equation}
  |\Psi\rangle = \text{C} {\mathcal A}
     [\Phi^{ \text{THSR}}({}^{10}\mathrm{Be})
     \phi_{11}(\mathbf{r}_{11})],
\end{equation}
where ${\mathcal A}$ is the antisymmetrizer and C is a normalization
constant. Then the density distribution $\rho(\mathbf{r}')$ of the last
valence neutrons is defined as
\begin{equation}
\label{eq:density}
  \rho(\mathbf{r}') = N_{c} \langle \Phi^{ \text{THSR}}(^{10}\mathrm{Be})
    \phi_{11}(\mathbf{r}_{11})|
    \delta(\mathbf{r}_{11} - \mathbf{X}_G - \mathbf{r}')
  | \Psi \rangle,
\end{equation}
where $N_{c}$ is the normalization constant \cite{Horiuchi1977}.  In
Fig.~\ref{fig:sigma-dens}, the sigma-orbit structure with a
distribution  falling into three different regions is observed in the
$y=0$ cross section, which is
consistent with pioneer works. The blue ellipse nodal surface is
clearly shown in this cross section. The last valence neutron is found
to have a very large distribution up to 8 fm in both $x$ and $z$
directions, which explains the giant RMS radius obtained for this
neutron. Comparing with pioneer works with AMD method, the
distribution in the central region is smaller in our
calculation. This comes from the larger central part of
nucleon-nucleon interaction in Eq.~\ref{eq:nn-central}, which leads to
a stronger $\alpha$-$\alpha$ attraction and closer $\alpha$-$\alpha$
distance. The neutron distribution is thus reduced because of the
Pauli blocking from two $\alpha$-clusters.

We use the angular momentum projection technique to extract the 
spectroscopic factor $S_{2s}$ from the ground state of ${}^{11}$Be.
This is obtained by calculating
the following squared overlap
\begin{equation}\label{s2s}
\begin{split}
S_{2s}&=N| \langle (\hat{P}^{0}_{0,0}\Psi^{\text{THSR} }
   ({}^{10}\text{Be})) \phi_{11}(\mathbf{r}_{11})| 
\hat{P}^{1/2}_{1/2,1/2}\Psi  \rangle|^{2}\\
  &=0.907.
\end{split}
\end{equation}
Here N is the coefficient originates from normalization
and many body effects. Physically, this means that the 
probability amplitude of the $|0^{+}\bigotimes 2s_{1/2}  \rangle$
occupation is about 91\% for the $1/2^{+}$ ground state of ${}^{11}$Be.
This value is almost the same as the value 93\% predicted theoretically by
Vinh Mau et al. in Ref.~\cite{Vinhmau1995}. Our result is larger than
the shell model predictions 74\% and 55\% in Refs.~\cite{Brown2001,Otsuka1993}.
This may partially due to the fact that our THSR wave function emphasizes
the long tail structure for valence neutron while shell model methods usually
deal with shrunk structure of nuclei. Recent experimental results
for the $S_{2s}$ are reported by Lima et al. in Ref.~\cite{Lima2007}.
The experimental values vary largely from 46\% to 87\%  with large 
error bars. Our result locates within error bars of four
experimental results but larger than others.

We also study how our new coupling strength $V_{ls}^{*}$
affects the spectroscopic factor $S_{2s}$. With a smaller 
$V_{ls}=$2000 MeV, the corresponding spectroscopic factor is $S_{2s}'$=0.933
for the $1/2^+$ ground state of ${}^{11}$Be. With our new coupling strength
$V_{ls}^{*}=$3200MeV, the spectroscopic factor is $S_{2s}$=0.907, 
which is 2.6\% smaller. This shows that our new $V_{ls}^{*}$ reduces 
the probability of the $|0^{+}\bigotimes 2s_{1/2} \rangle$ occupation
and increases the probability for the  $|2^{+} \bigotimes 1d_{5/2}  \rangle$ 
occupation in the $1/2^+$ ground state, which is also critical 
for the intruder ground state of ${}^{11}$Be \cite{Vinhmau1995}.

In summary, we performed a self-consistent calculation for the exotic
nucleus $^{11}$Be and other Be isotopes. The newly introduced valence
neutron creator provides a correct description for the nodal surface for
the $\sigma$-binding wave function in the ground state of $^{11}$Be.
Ground state energies and single-neutron separation energies, which
fit experimental results, are obtained for the $^{8-11}$Be isotopes by
systematic calculations with this new THSR wave function and
isospin-dependent spin-orbit interaction strength. The low energy
spectrum calculated { has the right ordering but remains slightly diluted. 
Explicit consideration of low lying states in ${}^{10}$Be may be necessary to improve.}
%also agrees with experimental values. 
However,
the intruder ground state $1/2^{+}$ with positive parity is correctly
reproduced what is not easy to achieve. The neutron halo structure observed from experiments is
illustrated with calculations of very large RMS radii and density 
distributions. The
giant spatial distribution of the last valence neutron is also
obtained, which agrees well with the phenomenological extraction from
experimental data. We also calculate and discuss the effects
of the spectroscopic factor $S_{2s}$ for the $1/2^+$ ground state of 
${}^{11}$Be.  This investigation {\ opens up} further evolution of the
nonlocalized clustering approach to the $\beta$-unstable region and
provides { very satisfying insights} for the exotic properties of the neutron
rich nucleus $^{11}$Be.

\begin{acknowledgments}
  This work is supported by the National Natural Science Foundation of
  China (grant nos 11535004, 11035001, 11375086, 11105079, 10735010,
  10975072, 11175085 and 11235001), by the National Major State
  Basic Research and Development of China (grant no 2016YFE0129300),
  by the Research Fund of Doctoral Point (RFDP), grant
  no. 20100091110028, and by the Science and Technology Development
  Fund of Macao under grant no. 068/2011/A. The authors would like to
  thank Professor Yanlin Ye, Professor Massaki Kimura, Dr. Hantao Li
   for valuable discussions.
\end{acknowledgments}

\end{document}